\documentstyle[12pt,fleqn]{article}  
\newcommand{\tr}{\; {\rm tr} }  
\newcommand{\Tr}{\mathop{\rm Tr}\nolimits}

\begin{document}  
\begin{titlepage}  
  
\begin{flushright}  
HU-EP-97/87  
\end{flushright}  
\bigskip  
  
\begin{center}  
\large\bf  
Generation of a QCD-induced Chern-Simons like term in QED   
\end{center}  
\vspace{0.5cm}  
\renewcommand{\thefootnote}{\fnsymbol{footnote}}  
\begin{center}  
D. Ebert\footnotemark[1],  
V.~Ch.~Zhukovsky\footnotemark[1]\footnotemark[2]\\ {\sl Institut  
f\"ur Physik, Humboldt--Universit\"at zu Berlin,\\  
Invalidenstra{\ss}e 110, D--10115 Berlin, Germany}  
\end{center}

\begin{abstract}  
After briefly reviewing the dynamical generation of a topological  
Chern-Simons (CS) term  in $QED_3$ the possibility of a  
Chern-Simons like term generation in the $QED_4$   
Lagrangian by quark loops interacting with a non-abelian background   
field of an $SU(2)$ model of $QCD$ is demonstrated.  
\end{abstract}  

\vspace{0.3cm}  
\setcounter{footnote}{1}  
\noindent\footnotetext{\noindent Supported in part by  
{\it Graduiertenkolleg ``Elementarteilchenphysik''} and {\it Scientific  
Exchange Program  between Moscow State University (Moscow) and  
Humboldt University (Berlin).}}  
\setcounter{footnote}{2}  
\noindent \footnotetext{\noindent On leave of absence from the  
Faculty of Physics, Department of Theoretical Physics, Moscow  
State University, 119899, Moscow, Russia.}  
\vfill  
\end{titlepage}

\renewcommand{\thefootnote}{\arabic{footnote}}  
\setcounter{footnote}{0}  
\setcounter{page}{1}  
  
\section{Introduction }  
  
In recent years much attention has been paid to the discussion of various   
nonperturbative effects in the theory of gauge fields, which are related  
to the role of topological terms and the vacuum condensate, as  
well as to the infrared behavior of the perturbation  
theory expansion at high temperature and in spaces of lower dimensions.   
  
Some of the  recently proposed QCD vacuum models are based upon   
investigations of  
the quark and gluon field interactions in the background of  
a certain model gluon condensate, i.e., a classical gauge  
field.  Along with  
instanton models of the QCD vacuum \cite{Shuryak}, efforts are undertaken   
to construct  
models, that are based upon various regular configurations of   
external non-abelian fields, the  
simplest of which is a constant chromomagnetic field of the abelian type,  
proposed by Savvidy \cite{savvidy}.  
As it was shown in \cite{Brown}, vector-potentials, by which homogeneous and   
constant fields are described, can only be of two types.  The first one is  
the so called  
covariantly-constant  field. In this case  
gluon fluctuations lead to a lowering of the vacuum energy   
\cite{savvidy}, while  the  vacuum  itself  turns  out  to be  unstable    
\cite{Nielsen}.  The  vector-potentials $G_\mu$ of the second type are  
gauge equivalent to constant non-commuting  
potentials, so that the field tensor $G_{\mu\nu}$ is defined through their  
commutator $iG_{\mu\nu}=[G_{\mu},G_{\nu}]$.  The  chromomagnetic  
field  configuration  formed by the non-abelian  potentials turned out to be  
stable with respect to the decay into pairs of real  particles  
\cite{Brown}, \cite{Agaev}.  In  
a number of publications \cite{Milsht}, \cite{Averin},  
modeling  the vacuum by the  
fields of the second type, a substantially  nonperturbative  
dependence of such objects, as the photon polarization operator, on the vacuum   
field was obtained.  
Active investigation of various radiative effects in non-abelian external  
fields has been initiated in this connection (see, e.g.\cite{[3_5]}).   
  
Since publication of the pioneer works \cite{deser}, where a topological mass term for gauge   
fields in (2+1)-dimensional spacetime was initially investigated, effects of topology produced by    
Chern-Simons modification of the gauge field Lagrangian attracted much attention.  
The study of  
radiative effects in 3-dimensional theories, carried out recently   
in Ref. \cite{Kostya}, \cite{Kolya}  
demonstrated  the importance  
of the Chern-Simons (CS) topological term for the regularization of  
infrared divergences in calculating higher loop diagrams even in the  
case when background fields are present.  
  
In particular, the one-loop electron mass   
operator and the photon  
polarization operator in (2+1)-QED in an external magnetic field at  
finite temperature and matter density \cite{Kostya}, as well as the mass   
operator of a quark in the (2+1)--dimensional QCD in an external   
chromomagnetic field \cite{Kolya} were calculated, and a nonanalytic   
dependence of the quantities   
studied on the external fields was demonstrated.   
  
A new discussion of the role of finite matter density in the effect  
of the dynamical Chern-Simons term generation in (2+1)--dimensional  
gauge theories has been initiated recently (see  
e.g. \cite{tseit,sissak}),  
showing that the problem of topological effects in gauge theories  
is far from being solved.  
  
A calculation of the photon polarization operator   
with exact consideration for the contribution of the non-abelian QCD   
vacuum condensate field  was  
carried out in the case of scalar quarks in  
\cite{Averin,mamed,mamsur}.  
   
Of special interest are possible Lorentz- and parity-violating modifications of (3+1)-dimensional   
electrodynamics,  investigated in \cite{jackiw}.   
These modifications are caused by adding to the $QED$ Lagrangian a Chern-Simons term that   
couples the dual electromagnetic tensor to a four-vector.  
In this paper  
we shall address ourselves to the possibility \cite{Buckow} that  
new nontrivial topological terms of the Chern-Simons type may be  
generated in (3+1)--dimensional $QED$, when induced by the fermion  
loops in the non-abelian vacuum condensate of $QCD$.  
  
  
\section{Chern-Simons term generation in an abelian background field  
in (2+1) dimensions}  
  
As it is well known \cite{deser},  spinor   
electrodynamics in (2+1)-dimensional space is described by the Lagrangian  
\begin{eqnarray}  
{\cal L}&=&{\cal L}_g + {\cal L}_f +{\cal L}_{int}  
\label{1}   
\end{eqnarray}  
with  
\begin{eqnarray}  
{\cal L}_g&=&-  
\frac{1}{4}F_{\mu\nu}F^{\mu\nu} +   
\frac{1}{4}\theta\varepsilon^{\mu\nu\alpha}F_{\mu\nu}A_\alpha  
\label{2}   
\end{eqnarray}  
for the gauge (electromagnetic) field, with  
$F_{\mu\nu}=\partial_{\mu}A_\nu - \partial_{\nu}A_\mu$,   
\begin{eqnarray}  
{\cal L}_f&=&\bar{\psi}(i{\gamma}\partial  - m)\psi  
\label{3}   
\end{eqnarray}  
for the fermion field, and  
\begin{eqnarray}  
{\cal L}_{int}&=&-j^{\mu}A_{\mu}  
\label{4}   
\end{eqnarray}  
for the interaction term, where   
\begin{eqnarray}  
j^\mu = -e\bar{\psi}\gamma^{\mu}{\psi}  
\label{5}   
\end{eqnarray}  
is the fermion current. The last term in ${\cal L}_g$ is the Chern-Simons  
term,   
\begin{eqnarray}  
{\cal L}_{CS}&=&\frac{1}{4}\theta\varepsilon^{\mu\nu\alpha}F_{\mu\nu}A_{\alpha},  
\label{6}   
\end{eqnarray}  
related to the CS secondary characteristic class. Here $\theta$ is the CS  
parameter which has a dimension of mass. It can be generated in the  
effective Lagrangian by quantum corrections \cite{deser,redlich}, which are affected by external  
fields, finite temperature and the density of medium (see e.g.  
\cite{tseit1,Kostya,sissak,tseit}, and references therein). In the 1-loop  
approximation this can be seen by calculating the photon polarization  
operator (PO)  
\begin{eqnarray}  
\Pi^{\mu\nu}(q)&=&ie^2 \int\frac{d^3 p}{(2\pi)^3} \tr { \left[ \gamma^{\mu}  
S(p+q/2)\gamma^{\nu}S(p-q/2) \right]},  
\label{7}   
\end{eqnarray}  
where   
\begin{eqnarray}  
S(p) = \frac{{\gamma}p+m}{p^2 - m^2}  
\label{8}   
\end{eqnarray}  
is the fermion Green's function operator. The polarization operator  
in $QED_3$ has two parts  
\begin{eqnarray}  
\Pi_{\mu\nu}&=&\Pi_{\mu\nu}^S + \Pi_{\mu\nu}^A ,  
\label{9}   
\end{eqnarray}  
where $\Pi_{\mu\nu}^S$ is symmetric and $\Pi_{\mu\nu}^A$ is  
antisymmetric.  In the (3+1)-dimensional  
case $\Pi_{\mu\nu}^A$ is absent, if there is no anisotropy in the space. It is  
evident that the presence of $\Pi_{\mu\nu}^A$ means the appearance of the  
vacuum anisotropy,  
induced by vacuum corrections.   
The so called dynamically induced CS term in PO has the  
following structure:  
\begin{eqnarray}  
\Pi_{\mu\nu}^{A}(q)&=&i\varepsilon_{\mu\nu\alpha}q^{\alpha}\Pi^{A}(q^2) .  
\label{10}   
\end{eqnarray}  
Here the value of $\Pi^{A}(q^2)$ at $q^2=0$ determines the induced CS topological  
photon mass $\theta_{ind} = \Pi^A(0)$ \cite{deser,redlich}.   
  
Note that in the presence of an external magnetic field $H$ the function   
$\Pi^{A}(k^2, H)$ in (10)  
becomes also a function of $H$ \cite{tseit1,tseit,sissak,Kostya}.

\section{Chern-Simons like term generation in a non--abelian  
background in (3+1) dimensions}  
   
In this section we shall demonstrate the appearance of an  
antisymmetric structure of the type (10) in (3+1)-$QED$ induced by the  
quark loop in a   
certain non-abelian background field of $QCD$.  
Let us consider a model with a massive quark   
field $\psi$, coupled both to an electromagnetic field $A_\mu$ and to an   
$SU(2)_C$  
gluon field $G _\mu$ in the (3+1)-dimensional space-time,  
which is described by the following Lagrangian   
\begin{eqnarray}  
{\cal L}&=&{\cal L}_g + {\cal L}_f +{\cal L}_{int}  
\label{11}  
\end{eqnarray}  
with  
\begin{eqnarray}  
{\cal L}_g&=&-\frac{1}{4}F_{\mu\nu}F^{\mu\nu}-  
\frac{1}{4g^2}G_{\mu\nu}^aG_a^{\mu\nu},  
\label{12}  
\end{eqnarray}  
\begin{eqnarray}  
{\cal L}_f&=&\bar{\psi}(i{\gamma}D - m)\psi ,  
\label{13}   
\end{eqnarray}  
and ${\cal L}_{int}$ is defined as in (4), (5).  
Here $G_{\mu \nu }^a=\partial _\mu G_\nu ^a-\partial _\nu G_\mu  
^a+f_{abc}G_\mu ^b G_\nu ^c$, and $D_\mu =\partial_\mu -i G_\mu$ ($ G_\mu =   
G^{a}_{\mu}   
T^a$, $T^a$ are   
the generators  
of the gauge group $SU(2)_C$).  
  
If a constant non-abelian rotationally-invariant background field  
$G_\mu = const$ of the   
second type (see Introduction) is present,  
then, neglecting contributions of the  gluon  
fluctuations around this field, we may calculate the 1-loop PO of a  
photon ($P_\mu = p_\mu + G_\mu$)  
\begin{eqnarray}  
\Pi^{\mu\nu}(q)&=&ie^2 \int\frac{d^4 p}{(2\pi)^4} \Tr { \left[ \gamma^{\mu}  
S(P+q/2)\gamma^{\nu}S(P-q/2)\right]},  
\label{15}   
\end{eqnarray}   
moving through the QCD vacuum, and interacting with spin $1/2$-quarks in the  
loop. Here the quark Green's function is given by the following  
expression with exact consideration for the vacuum field $G_\mu$  
\begin{eqnarray}  
S(p) = \frac{1}{{\gamma}P-m}=\frac{1}{P^2 - m^2 -  
\frac{1}{2}\sigma G}({\gamma}P+m) ,  
\label{16}   
\end{eqnarray}  
where $\sigma G = \sigma_{\mu\nu}G^{\mu\nu}$.   
Taking into account, that the antisymmetric  
term in PO appears as a structure which is proportional to an antisymmetric  
tensor, we expand (15) up to a linear in $\frac{1}{2}{\sigma}G$ term.  
For the antisymmetric part of PO  
\begin{eqnarray}  
\Pi_{\mu\nu}^A&=&\frac{1}{2}(\Pi_{\mu\nu} - \Pi_{\nu\mu}) ,  
\label{17}   
\end{eqnarray}  
we then obtain  
\begin{eqnarray}  
\Pi_{\mu\nu}^A &=&ie^2 \int \frac{d^4p}{(2\pi)^4} \Tr \left[  
(\frac{1}{2}{\sigma}G)\pi_{\mu\nu}\frac{1}{(P+q/2)^2 - m^2} - (\mu  
\leftrightarrow \nu )   
\right].  
\label{18}   
\end{eqnarray}  
Here  
\begin{eqnarray}  
\pi^{\mu\nu}(P,q)&=&\left[ \gamma^{\mu}  
S(P+q/2)\gamma^{\nu}S(P-q/2)\right]_{{\sigma}G=0}.  
\label{19}   
\end{eqnarray}  
The integral in (\ref{18}) is UV finite, and after further expansion   
in (\ref{18}), (\ref{19}) up to terms linear in $G_\mu$ we obtain   
$\Pi_{\mu\nu}^A$ which is nonvanishing at $q^2\rightarrow 0$   
\begin{eqnarray}  
\Pi_{\mu\nu}^A(q, G)&=&\frac{5}{6\pi^2}\frac{e^2}{m^2} \tr   
{(G_{\mu}G_{\nu}(q_{\alpha}G^\alpha))}.  
\label{20}   
\end{eqnarray}  
For the rotationally-symmetric background field  
\begin{eqnarray}  
G_1^1&=&G_2^2=G_3^3=\sqrt{\lambda}, G_0^a=0, \quad H_i^a=\delta _i^a \lambda  
(i=1,2,3),   
\label{14}   
\end{eqnarray}  
we have:  
\begin{eqnarray}  
\Pi_{\mu\nu}^A(q, \lambda)&=&\frac{5i}{24\pi^2} \frac{e^2}{m^2}\lambda^\frac{3}{2}  
\varepsilon_{\mu\nu\alpha} q^\alpha ,  
\label{21}   
\end{eqnarray}  
(where now $\mu, \nu, \alpha = 1, 2, 3$), which is very similar in  
structure to the 3d-result (10). The photon topological mass is equal to  
\begin{eqnarray}  
\theta_{ind}(\lambda)&=&\frac{5}{24\pi^2} \frac{e^2}{m^2}\lambda^\frac{3}{2}.  
\label{22}   
\end{eqnarray}  
We see that, like in a (2+1)-dimensional case, $QCD$ vacuum corrections may   
induce an anisotropy in (3+1)-$QED$, if a non-abelian constant  
background field is present.  
  
In the more general case of strong background fields of large intensity  
$\lambda$ the induced antisymmetric structure of $QED$ should have the  
form ($\mu, \nu, \alpha = 1, 2, 3$):  
\begin{eqnarray}  
\Pi_{\mu\nu}^{A}&=&i\varepsilon_{\mu\nu\alpha}q^{\alpha}\Pi^{A}(q^2, \lambda) ,  
\label{23}   
\end{eqnarray}  
with  
\begin{eqnarray}  
\Pi^{A}(0, \lambda)&=&\theta_{ind}(\lambda).  
\label{24}   
\end{eqnarray}  
Using one of the eigenvectors of PO in the external field $G_\mu$:  
\begin{eqnarray}  
s_\beta  
&=&-\frac{2i}{3}\lambda^{-\frac{3}{2}}\varepsilon_{\beta\sigma\lambda\rho}   
\tr{(G^\sigma G^\lambda G^\rho)}  
\label{25}   
\end{eqnarray}  
we can write the corresponding CS structure in the $QED$ action in a  
4-dimensional covariant way  
\begin{eqnarray}  
S_{CS}&=&\frac{1}{4}\int  
d^4x\theta_{ind}(\lambda)\varepsilon^{\mu\nu\alpha\beta}F_{\mu\nu}A_{\alpha}s_\beta  
.  
\label{26}   
\end{eqnarray}  
One can see that the topological current vector \cite{deser},   
introduced for an arbitrary varying field $G_{\mu}(x)$,     
\begin{eqnarray}  
X^\beta &=&4\varepsilon^{\beta\sigma\lambda\rho} \tr {(G_\sigma  
\partial_\lambda G_\rho - \frac{2i}{3} G_\sigma G_\lambda G_\rho)}  
\label{27}   
\end{eqnarray}  
is proportional to $s_\beta$ (25) in the case   
of a constant field. Thus one might expect that (26) after replacement   
$s^{\beta} \rightarrow X^{\beta}$ holds also for an  
arbitrary varying background field with non-zero topological charge.  
  
\section{Discussion}  
  
The induced CS structure discussed above may provide one of the physical  
mechanisms for  possible unusual phenomena in the propagation of light through   
the universe. For instance, evidence for an   
anisotropy in electromagnetic field propagation over cosmological  
distances was reported recently in \cite{ralston}. They claimed to observe a new effect  
(extracted from Faraday rotation) of rotation  of the  
plane of polarization of electromagnetic radiation propagating over  
cosmological distances, correlated with the angular position and  
distance of the source. It should be pointed out that this  observation of anisotropy initiated   
many controversial comments concerning possible interpretation of the results of \cite{ralston}   
(see e.g. \cite {carroll}, and a more recent   
publication \cite {ralston1} and references therein). Such an evidence, in case it is confirmed,   
might signal for a non-trivial topological  
structure of the space over large scales, explained by microscopic  
reasons.  
The authors of \cite{ralston} discussed an effective action  
with a gauge invariant coupling of $A_{\mu}$ and $F_{\mu \nu}$ to a   
certain background vector $s_\mu$ of the form (26). The possibility that such a term in (3+1)-  
$QED$ Lagrangian could produce an  
anisotropy in electromagnetic field propagation was pointed out first by Carroll, Field  
and Jackiw \cite {jackiw}.  
One might relate $s_\mu$ to some intrinsic ``spin axis''  
of an anisotropic universe, associated with axion   
type domain walls  or some other specific condensate  
structures of the type considered above.   
  
As it is well known, there exists a number of nontrivial topological  
solutions of field equations, such as, e.g., instantons.  The   
non-abelian constant  
field case, discussed above, provides just a simple idea of a  
possible birefringence effect in the propagation of electromagnetic  or   
other gauge fields that may arise under the influence of a nontrivial   
topological background. More realistic examples are to be discussed in   
subsequent publications.

\section*{Acknowledgments}  
The authors are grateful to R.Jackiw for useful comments. \\  
One of the authors (V.Ch.Zh.) gratefully acknowledges the  
hospitality of Prof. Mueller-Preussker and his colleagues at the  
particle theory group of the Humboldt University extended to him  
during his stay there.

\end{document}